\documentclass[superscriptaddress,twocolumn,bibnotes, amsmath,amssymb,aps,pra,floatfix]{revtex4-2} 

\usepackage{graphicx}
\usepackage{dcolumn}
\usepackage{bm}
\usepackage{hyperref}
\usepackage{physics}

\usepackage{braket}
\usepackage{multirow} 
\usepackage{xcolor}
\usepackage{lipsum} 

\newcommand{\be}{\begin{equation}} 
\newcommand{\ee}{\end{equation}} 

\begin{document}

\title{\texorpdfstring{Emergent $s$-wave interactions in orbitally active quasi-two-dimensional Fermi gases}{}} 

\newcommand{\Toronto}{Department of Physics and CQIQC, University of Toronto, Ontario M5S~1A7, Canada}
\newcommand{\HKU}{Department of Physics and Hong Kong Institute of Quantum Science and Technology, The University of Hong Kong, Hong Kong, China}
\newcommand{\Trento}{Pitaevskii BEC Center CNR-INO and Dipartmento di Fisica Universit\`a di Trento}

\author{C. J. Dale$^{1,*}$,
K. G. S. Xie$^{1,*}$,
K. Pond Grehan$^{1}$,
Shizhong Zhang$^{2}$,
J. Maki$^{3}$,
J. H. Thywissen$^{1}$ \\~\\
\small \textit{$^{1}$Department of Physics and CQIQC, University of Toronto, Ontario M5S~1A7, Canada}\\
\small \textit{$^{2}$Department of Physics and Hong Kong Institute of Quantum Science and Technology, The University of Hong Kong, Hong Kong, China}\\
\small \textit{$^{3}$Pitaevskii BEC Center CNR-INO and Dipartmento di Fisica Universit\`a di Trento}\\
\small $^{*}$These authors contributed equally to this work.\\\vspace{5pt}
}

\date{\today}

\begin{abstract} 
We investigate the scattering properties and bound states of a quasi-two-dimensional (q2D) spin-polarized Fermi gas near a $p$-wave Feshbach resonance. Strong confinement promotes the out-of-plane spatial wave functions to a discrete, gapped orbital degree of freedom. Exchange-antisymmetric orbital pair wave functions are predicted to give rise to low-energy q2D interactions with $s$-wave symmetry. Using radiofrequency (rf) spectroscopy, we observe the signature power-law scaling and the dimensional-crossover feature anticipated for the emergent $s$-wave channel. Additionally, we demonstrate that two types of low-energy dimers, with either $s$-wave and $p$-wave symmetry, could be formed via rf spin-flip association from an orbital mixture. These findings illustrate how gapped orbital degrees of freedom can provide additional control over scattering symmetries in strongly confined ultracold gases. 
\end{abstract}
\maketitle

{\em Introduction.}
%
Orbital degrees of freedom are natural in electronic materials, reflecting the valence structure of their atomic constituents. Conversely, neutral atoms in optical lattices typically occupy the lowest-energy band and interact through $s$-wave interactions. Pioneering experiments employing excited orbital bands have revealed new phenomena \cite{Muller:2007,Bakr:2011,Schmiedmayer:2021,Hartke:2022,Lee:2023} including chiral many-body states \cite{Hachmann:2021}. Furthermore, theoretical work suggests new types of quantum simulation \cite{Dutta2015,Li:2016,Mamaev2020,mamaevpwave2021} and rich physics of extended Hubbard models \cite{Liu:2008,ZhangZ:2012,Fedorov:2017}. Another avenue of investigation employs interactions with higher angular-momentum partial waves, such as $p$-wave scattering~\cite{Regal:2003,Suno2003PRL,Ticknor:2004,Zhang:2004,Schunck:2005,Gunter:2005,chevy:2005}, in optical lattices. Due to the anisotropic nature and nodal structure of $p$-wave scattering, interatomic interactions depend crucially on the orbital bands in the lattice, leading to novel scattering modalities. 

Of particular interest is the emergent even-wave scattering channel recently observed in a spinless Fermi gas confined in a quasi-one-dimensional (q1D) geometry, triggered by the population of excited orbitals~\cite{Jackson:2023}. Here we demonstrate that this phenomenon also occurs in quasi-two-dimensional (q2D) systems. Unlike previous studies with spin-polarized fermions in q2D, which only populated the ground state of the confining potential perpendicular to the 2D plane~\cite{Gunter:2005,Waseem:2016,Waseem:2017, Zhang:2017, Kurlov:2017}, 
we intentionally excite motional states out of the q2D plane. We show that activating this orbital degree of freedom leads to scattering states and bound states with an emergent $s$-wave symmetry within the q2D subspace.

The emergence of an $s$-wave scattering symmetry in a spinless Fermi gas can be understood as follows. The strong q2D confinement potential breaks the full rotational symmetry of free space into a reflection symmetry about the plane and rotation within the plane. Due to the reflection symmetry, the two-body wave function must have definite parity in the direction perpendicular to the q2D plane. In the case of even orbital parity, it is necessary that scattering have {\em odd} exchange symmetry within the q2D plane. This leads to the standard $p$-wave scattering in 2D. In terms of single-particle occupation, this first scenario is realized when both atoms occupy the lowest orbital state of the out-of-plane direction, which is typically the case considered in the literature. On the other hand, if the out-of-plane parity is odd, then necessarily scattering within the q2D plane must have {\em even} exchange symmetry. This leads to emergent $s$-wave scattering in the plane dominating at low energy. In terms of single-particle occupation, this second scenario corresponds to one atom in the ground orbital and one atom in the excited orbital; once symmetrized, this pair state is an orbital singlet. 

The structure of the Letter is as follows. 
First, we characterize the short-range symmetry of atom-atom interactions by the high-frequency scaling of their radiofrequency (rf) association rate. When the excited orbital band is populated, we observe $s$-wave scattering in the q2D continuum. 
Second, we measure the pair binding energy in q2D. We observe pairs with both $s$- and $p$-wave symmetry in the 2D plane. Both energies agree with recent theoretical predictions \cite{companion}. 

{\em Sample preparation.} 
Our experiments begin with an ultracold Fermi gas of $N_\mathrm{tot} = 8(1) \times 10^4$ potassium-40 ($^{40}$K) atoms cut into isolated planar traps by an optical standing wave. The out-of-plane confining potential is approximately harmonic with angular frequency $\omega_\perp$, giving rise to an energy gap $\hbar \omega_\perp$ between the ground and first excited orbital. When $\hbar \omega_\perp$ far exceeds the in-plane Fermi energy and temperature, each planar sample is energetically restricted to a q2D subspace. At low energies, the asymptotic scattering wave function in the weakly confined directions resembles that of a true two-dimensional system \cite{Petrov:2001,LevinsenParish:2015}.

\begin{figure*}[tb!] \centering
\includegraphics[width=7in]{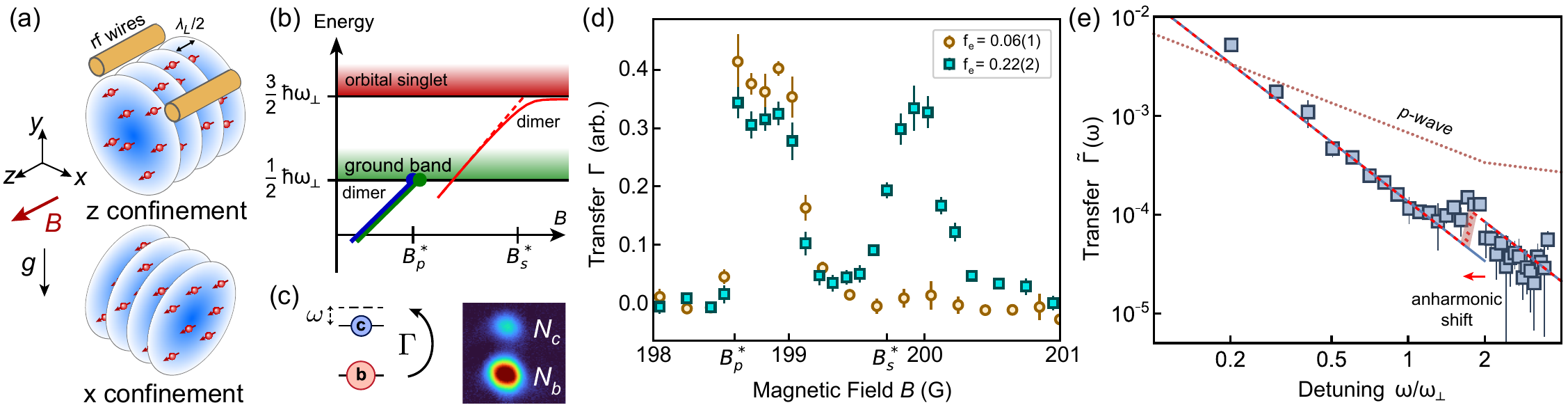}
\caption{
{\em Radiofrequency spectroscopy of scattering in the continua.} 
(a) Two confinement geometries are used: a standing wave along $z$ or $x$ creates $xy$ or $yz$ planar systems. The $z$-confinement scenario is cylindrically symmetric since the Feshbach field is oriented perpendicular to the plane. 
(b) For $bb$ atom pairs in the orbital ground state, a continuum of scattering states exist above a threshold energy set by the zero-point energy of $\frac12 \hbar \omega_\perp$ in the strongly confined direction. Below threshold, there are discrete $p$-wave dimer states (shown as blue and green lines, representing the near degeneracy of the $M_L=\pm1$ states in the $z$-confinement geometry). For $bb$ atom pairs with an additional quantum of orbital motion, the scattering continuum exists above a threshold of $\frac32 \hbar \omega_\perp$. A discrete $s$-wave dimer state exists below threshold (red line).
While q2D scattering in the ground band has $p$-wave symmetry, scattering in the orbitally excited continuum has $s$-wave symmetry. These q2D $s$-wave interactions are Feshbach-enhanced around $B^*_s$, where the dimer state is magnetically tuned close to continuum.
(c) An illustration of the rf transfer $\Gamma$ from state $\ket{b}$ to $\ket{c}$ with detuning $\omega$. An example absorption image of the two spin states after time-of-flight is shown.
(d) Transfer rate $\Gamma$ versus magnetic field $B$ with confinement of $V_L=40\,E_R$ applied in the $z$ direction. The spectroscopic pulse is $50$\,kHz detuned from the bare $\ket{b}$-to-$\ket{c}$ resonance. As described in the text, the increased $\Gamma$ just above $B^*_p$ and $B^*_s$ indicate increased correlation strengths. Measurements at two $f_e$ values demonstrate that the $s$-wave correlations require an orbital mixture. 
(e) Dimensionless rf transfer rate $\widetilde\Gamma$ versus detuning $\omega$, for an ensemble prepared with $f_e = 0.29(2)$ at $B=B_*^s+0.42(1)$\,G with $V_L=80\,E_R$. The signature $s$-wave scaling $\widetilde\Gamma \propto \omega^{-2}$ is observed [see Eq.~\eqref{eq:Gamma_q2d_s}]. An additional term in the transfer rate activates at $\omega \gtrsim 2 \omega_{\perp}$, resulting in an increase in $\widetilde\Gamma$. The $p$-wave transfer equation [Eq.~\eqref{eq:Gamma_q2d_p}] is plotted as a dotted line for comparison. An anharmonic correction is applied to the red dashed line, with the band near the $1.6 \omega_{\perp}$ feature representing lattice depth uncertainty. 
}
\label{fig:s-wave_scattering}
\end{figure*}

An external magnetic field is applied to tune the interactions with the $p$-wave Feshbach resonance at magnetic field $B=198$\,G~\cite{Regal:2003,Gunter:2005}. The low-energy scattering amplitude for such an interaction in free space defines the 3D $p$-wave scattering volume $V_\mathrm{3D}$ and effective range $R_\mathrm{3D}>0$. For $^{40}$K, the magnetic-field dependence of $V_\mathrm{3D}$ and $R_\mathrm{3D}$ has been determined by comparing coupled-channels calculations to experimental data \cite{Ticknor:2004,AhmedBraun:2021}. The scattering parameters also depend on the projection $M_L$ of the $L=1$ orbital angular momentum onto the magnetic field axis. Feshbach resonances for $M_L=0$ and $|M_L|=1$ exist near $198$\,G, between atoms in the second-lowest Zeeman state $\ket{b}$, adiabatically connected to the low-field eigenstate $\ket{F=9/2, m_F=-7/2}$. 

Figure~\ref{fig:s-wave_scattering}(a) shows the two different confinement geometries used: one where the lattice potential is parallel to the external magnetic field (``$z$ confinement''), and one where the potential is perpendicular (``$x$ confinement''). A crossed optical dipole trap (ODT) imposes the in-plane trapping, with frequencies $[(\omega_x, \omega_y), (\omega_y, \omega_z)] = 2\pi\,[(170, 440),(440, 440)]\,$Hz for parallel and perpendicular magnetic field geometries, respectively.
The lattice depth $V_L$ of the $\lambda_L=760.6$\,nm standing wave is slowly ramped to a shallow potential $\sim 4\,E_R$, then quickly ramped to the desired set point, where $E_R = h^2/2m\lambda_L^2 \approx h \times 8.6\,$kHz is the recoil energy, and $m$ is the atomic mass. 

The atom number and temperature distributions in the ensemble of planar traps are estimated with a loading model, where the transfer between the gas in the ODT and the array of planar traps is assumed to be isentropic. The chemical potential and temperature of the planar traps are found from imposing entropy and number conservation between the ODT and the planes. Transport between planes is suppressed by the final depth of the lattice potential after it is ramped beyond $10\,E_R$. Typical atom numbers in the q2D planes are estimated to be $3200$ and $1200$, for the $z$- and $x$-confinement geometries, respectively, with a typical reduced temperature $k_B T/E_F = 0.4$, where $E_F$ is the in-plane Fermi energy, typically $h \times 22\,$kHz. 

We define reference magnetic fields, $B^*_p$ and $B^*_s$, to be where the bare 3D $p$-wave dimer crosses the confinement-shifted continua [see Fig.~\ref{fig:s-wave_scattering}(b)].  
For q2D $p$-waves, the threshold energy is the zero-point energy of the confinement, such that $B^*_p$ is given by $E_\mathrm{3D}(B^*_p) = \frac12 \hbar \omega_\perp$, 
where $E_\mathrm{3D} = -\hbar^2 R_\mathrm{3D}/(m V_\mathrm{3D})$ for $V_\mathrm{3D} < 0$. 
For q2D $s$-waves, the continuum requires an additional quantum of orbital excitation, such that $B^*_s$ is given by 
$E_\mathrm{3D}(B^*_s) = \frac32 \hbar\omega_\perp$. 

{\em Emergent $s$-wave scattering.}  
The nature of interactions is probed by measuring the rf transfer rate $\Gamma$ from the interacting spin state $\ket{b}$ to a weakly interacting final spin state $\ket{c}$ [see Fig.~\ref{fig:s-wave_scattering}(c)].
Here $\omega = \omega_\mathrm{rf} - \omega_{bc}$, the detuning from the single-particle resonance. When $\hbar \omega \gg E_F$, $\Gamma$ is proportional to 
contact parameters \cite{Tan:2008a,Tan:2008b,Tan:2008c,Braaten:2008tc,Braaten:2008ez,Werner:2009vb,Zhang:2009kq,Braaten2010,Braaten:2012gh,Werner:2012um,Werner:2012hy,Yoshida:2015hh,Yu:2015go,Luciuk:2016gr,He:2016bn,HuiHu:2016}, as has been explored near $p$-wave interactions in 3D and 1D \cite{Luciuk:2016gr, Jackson:2023} and for $s$-wave interactions in 2D \cite{Frohlich:2012,Luciuk:2017}. 
A typical time sequence transfers the gas to the interacting state $\ket{b}$ with a resonant rf pulse in 40\,\textmu s, then held for 160\,\textmu s for correlations to develop, before a spectroscopic pulse transfers a perturbative fraction of the ensemble to state $\ket{c}$. Spin populations are then measured in state-selective time-of-flight imaging. 
The rf transfer rate $\Gamma = N_c/t_{\text{rf}}$, where $N_c$ is the number of transferred $\ket{c}$ atoms and $t_{\text{rf}}$ is the time of the spectroscopy pulse,  
is rescaled to a dimensionless rate as
$\widetilde{\Gamma} = E_F \Gamma /(\pi \hbar \Omega^2 N_\mathrm{tot})$, 
where $\Omega$ is the Rabi frequency. 

The orbital populations are controlled by modulating the lattice potential and through optimized ramp-on of the optical lattice, as in prior work \cite{Jackson:2023}. The ratio of atoms loaded into the ground and excited bands is measured by bandmapping. 
Fig.~\ref{fig:s-wave_scattering}(d) shows the transfer rate over a range of magnetic fields for $\omega = 2 \pi \times 50$\,kHz when the lattice is parallel to the magnetic field. Measurements for two different lattice loading procedures are shown, one with reduced first-excited-band population $f_e=6\%$ (circles), and one with increased population $f_e = 22\%$ (squares). We see that the peak at $B \gtrsim B^*_s$ appears only for ensembles loaded with significant $f_e$, whereas the peak at $B \gtrsim B^*_p$ appears in either case. This behavior is observed for both confinement directions, and confirms the identification of the peak above $B^*_s$ as an inter-band interaction. 

Figure~\ref{fig:s-wave_scattering}(e) shows an rf spectrum taken at fixed magnetic field just above $B^*_s$. For the high-frequency regime probed here, the predicted transfer rate \cite{companion} is 
\be \label{eq:Gamma_q2d_s}
\Gamma(\omega) \to \frac{\Omega^2 C_s}{2}\left[\frac{1}{\omega^2}+\frac{3\theta(\omega-2\omega_\perp)}{2 \omega^2}\right]\,, \ee
for $E_F/\hbar \ll \omega < 4 \omega_\perp$. Here $C_s$ is the contact parameter that is the thermodynamic conjugate of the q2D $s$-wave scattering length, $a_s$. This scattering length emerges from the short-range $p$-wave parameters \cite{companion}. In contrast, the q2D $p$-wave transfer rate would be, for $E_F/\hbar \ll \omega < 4 \omega_\perp$, 
\be \label{eq:Gamma_q2d_p} 
\Gamma(\omega) \to \frac{\pi \Omega^2 C_p}{2}\left[\frac{1}{\omega} +\frac{(\omega-2\omega_\perp)\theta (\omega-2\omega_\perp)}{2\omega^2}\right]\,,
\ee
where $C_p$ is the contact parameter that is the thermodynamic conjugate of the q2D $p$-wave scattering area, $a_p$. 
These $\Gamma(\omega)$ are both shown as lines in Fig.~\ref{fig:s-wave_scattering}(e). Qualitatively, one sees better agreement with Eq.~\eqref{eq:Gamma_q2d_s}. 
A quantitative comparison can be made by a free power-law fit for rf detunings $0.5\,\omega_\perp < \omega < 1.6\,\omega_\perp$, giving $-2.16(11)$, which supports the leading $\omega^{-2}$ scaling that is a signature of q2D $s$-wave scattering. 
This is evidence that, as discussed above, the exchange-antisymmetric (singlet) orbital wave function in the out-of-plane direction enables spin-polarized fermions to scatter with $s$-wave symmetry in the q2D plane.

\begin{figure*}[bt!]
\centering
\includegraphics[width=6.9in]{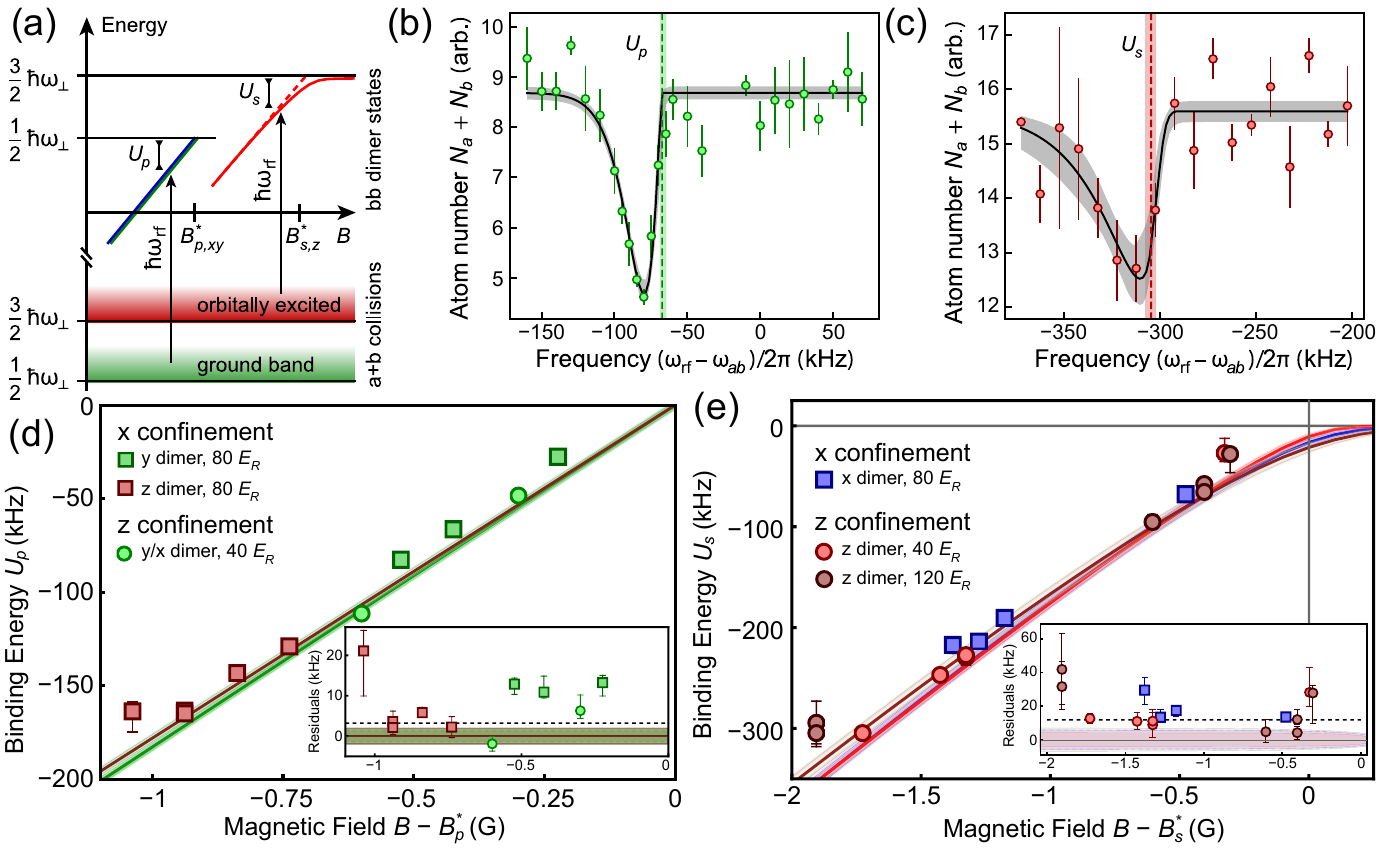}
\caption{{\em Dimer binding energies.}
(a) Dimers are associated by rf spectroscopy from an $ab$ spin mixture. In the $z$-confinement geometry depicted here, the $p_x$ and $p_y$ dimer states are degenerate. The binding energy ($U_p$ or $U_s$) places rf resonances below the $\ket{a}$-to-$\ket{b}$ single-atom spin-flip frequency. 
(b) Example of a spectroscopy measurement of a $p$-wave dimer in the $x$-confinement geometry, with $V_L=80\,E_R$ and $B=198.3$\,G. Vertical error bars are statistical. The black line shows a lineshape function (see main text), yielding a best-fit $U_{p}$, indicated with the vertical line.  
(c) $s$-wave dimer measurement in the $z$-confinement geometry, with $V_L=40\,E_R$ and $B=198.0$\,G. The best-fit $U_{s}$ is indicated the vertical line. 
(d) Summary of measured $p$-wave binding energies $U_s$ at various magnetic fields. Square and circular data indicate $z$- and $x$-confinement geometries, respectively. 
Solid lines show q2D dimer energies Eq.~\eqref{eq:2DpwaveDimer}, with no adjustable parameters. Residual differences are shown in the inset. 
(e) Summary of best-fit $s$-wave binding energies, $U_s$, versus magnetic field, for several confinement geometries and lattice depths, as labeled. 
Solid lines are solutions to the exact q2D binding energy theory given by Eq.~\eqref{eq:2DswaveDimer}, with no free parameters. Residuals differences are shown in the inset. 
Vertical error bars denote fit-parameter uncertainty. Shaded bands represent systematic uncertainties. The anharmonic correction is shown as a dashed line in both residual plots. 
\label{fig:dimer}}
\end{figure*}

Beyond its leading power-law behaviour, the rf transfer spectrum in Fig.~\ref{fig:s-wave_scattering}(e) shows a non-monotonic feature at $\sim 1.6\omega_\perp$, with qualitative similarity to the jump at $2\omega_\perp$ anticipated by Eq.~\eqref{eq:Gamma_q2d_s}. The origin of this feature is projection of the interacting $bb$ pair state onto the $\frac72 \hbar \omega_\perp$ continuum in the $bc$ channel. This exit channel only becomes energetically open above an orbital gap that is $2 \hbar\omega_\perp$ in the harmonic limit. Its presence is also a signature that strong interactions admix the next odd-parity oscillator state (at $\frac72 \hbar \omega_\perp$) into the $bb$ interacting pair wave function. In q1D, the singular density of states promotes this step into a singularity \cite{Jackson:2023}. Although an even-parity oscillator state (at $\frac52 \hbar \omega_\perp$) also appears as a second term in the q2D $p$-wave case, Eq.~\eqref{eq:Gamma_q2d_p}, $\Gamma$ remains monotonically decreasing, with a change only in the $\omega^{-1}$ amplitude. 
The dashed line shows Eq.~\eqref{eq:Gamma_q2d_s} including lattice anharmonicity \cite{SM} and inhomogeneity, shifting the position to $\sim 1.6 \,\omega_\perp$ and smoothing the step, which are both more consistent with the observed transfer rate. 

{\em q2D bound states.} 
A complementary characterization of pairwise interaction is through the bound states of the scattering potential. 
As shown in Fig.~\ref{fig:s-wave_scattering}(b), $p$-wave dimer states are expected below the ground band, and an $s$-wave dimer is expected below the orbitally excited continuum, both for $bb$ atom pairs. 

Our approach to measuring bound-state energies is shown in Fig.~\ref{fig:dimer}(a). A spin mixture of $\ket{a}$ and $\ket{b}$ atoms is prepared with a controlled $f_e$. 
Dimers are associated with an rf pulse that is detuned by $\omega_\mathrm{rf} - \omega_{ab}$ from the bare $\ket{a}$-to-$\ket{b}$ resonance.
For the (higher-symmetry) $z$-confinement scenario depicted in Fig.~\ref{fig:dimer}(a), the q2D $p$-wave dimer state is two-fold degenerate, hybridizing with either the $p_x$ or $p_y$ 3D dimer states; the q2D $s$-wave dimer state hybridizes with the $p_z$ 3D dimer state. Since collisions in the $ab$ channel occur in one of two bands offset by $\hbar \omega_\perp$, the differing confinement energies do not affect the rf frequency: instead, a resonance is expected near $\omega=U_s/\hbar$ or $\omega = U_p/\hbar$, offset by the incident kinetic energy %
\footnote{For the $x$-confinement scenario, not depicted in Fig.~\ref{fig:dimer}(a), dipole-dipole splitting lifts the degeneracy of the $p$-wave dimers -- now corresponding to $y$ and $z$ directions -- leading to three distinct dimer energies.}. Here $U_s$ and $U_p$ are the binding energies of the effective q2D dimers. 

Figure~\ref{fig:dimer}(b) shows an example rf spectrum in the $x$-confinement geometry. The vertical axis is the total atom number measured in $\ket{a}$ and $\ket{b}$ by Stern-Gerlach separation and absorption imaging after release from the trap. The lifetime of the $bb$ dimers is assumed to be short compared to the pulse and hold times, so that atom loss is a faithful signature of dimer association. This spectrum shows an asymmetric loss features, which we identify a $p$-wave dimer by the resonant frequency. An example rf spectrum with an $s$-wave dimer loss feature in the $z$-confinement geometry is shown in Fig.~\ref{fig:dimer}(c).
%

Spectra are fit with a model that includes an energy-dependent Franck-Condon factor, a thermal distribution of $ab$ collisional kinetic energy, and confinement inhomogeneity \cite{SM}. 
The $p$-wave dimer shown has a right edge fit as the binding energy $U_p/h= -66\substack{+4 \\ -1}\,$kHz. The $s$-wave dimer loss feature is broader, with the binding energy $U_s/h= -305(3)$\,kHz found near the maximal transfer frequency. 

Dimer spectra are taken at lattice depths of $40$, $80$ and $120$\,$E_R$. Typically, $35\%$ of atoms are prepared in the orbitally excited band. 
Similar rf spectra were obtained for both confinement geometries. Figures~\ref{fig:dimer}(d) and \ref{fig:dimer}(e) summarize the best-fit $U_p$ and $U_s$ as a function of $B-B^*_p$ and $B-B^*_s$, respectively. Measurements are compared to the predicted dimer energies \cite{companion}  determined by solving for $\mathcal{E}_p = U_p / (2 \hbar \omega_\perp)$ and $\mathcal{E}_s = U_s / (2 \hbar \omega_\perp)$ according to  
\begin{widetext} \begin{align} \label{eq:2DpwaveDimer} 
\frac{-\hbar R_\mathrm{3D}}{\omega_\perp m V_\mathrm{3D}} & = \frac12 - 2\mathcal{E}_p + \frac{3}{ \pi^{1/2}} \frac{R_\mathrm{3D}}{a_\perp} \int_0^\infty \!\! du  \left[ \frac{u e^{-\mathcal{E}_p u} }{u + 4 \lambda} \frac{(1-\lambda)^2 e^{-u} + 2 \mathcal{E}_p \eta}{ 8\lambda \eta^{3/2}} - \frac{1}{(u+ 4 \lambda)^{5/2}} + \frac{2\mathcal{E}_p - 1/2}{3(u+4 \lambda)^{3/2}} \right] \, , \\
\frac{- \hbar R_\mathrm{3D}}{\omega_\perp m V_\mathrm{3D}} &  = \frac32 - 2\mathcal{E}_s + \frac{3}{ \pi^{1/2}} \frac{R_\mathrm{3D}}{a_\perp} \int_0^\infty \!\! du \left[ \frac{e^{-\mathcal{E}_s u} }{u + 4 \lambda} \frac{1}{\eta^{3/2}} - \frac{1}{(u+ 4 \lambda)^{5/2}} + \frac{2\mathcal{E}_s - 3/2}{3(u+4 \lambda)^{3/2}} \right]\,, \label{eq:2DswaveDimer} 
\end{align} \end{widetext}
where $\eta=(1+\lambda)^2 - e^{-u}(1-\lambda)^2$, $\lambda=2(\Lambda a_\perp)^{-2}$, and $\Lambda$ is a cutoff parameter, here taken to be $R_\mathrm{3D}^{-1}$, such that $\lambda \sim 10^{-2}$. These relations describe the location of the pole of the two-body T-matrices in q2D. The integrals in each expression give the contribution to the bound state energy of virtual excitations in the q2D geometry, as well as the renormalization of the 3D scattering parameters \cite{companion}. The solutions to Eqs.~\eqref{eq:2DpwaveDimer} and \eqref{eq:2DswaveDimer} are shown in Figs.~\ref{fig:dimer}(d) and \ref{fig:dimer}(e); shaded bands around theory curves are a consequence in systematic uncertainty in $V_L$. 

The differences between measurements and theory are shown in the residual plots, inset to Figs.~\ref{fig:dimer}(d) and \ref{fig:dimer}(e). In both cases, dimer states are observed to be slightly less bound than predicted. Two possible reasons for this trend are interaction effects in the lineshape model and anharmonic corrections to the transition energies. As discussed in \cite{SM}, we estimate the anharmonic correction to be $+3$\,kHz in the $p$-wave case and $+12$\,kHz in the $s$-wave case (see dashed lines in insets). Overall, since there are no free parameters in this comparison, the agreement is satisfying.

{\em Conclusion.}
We have observed $s$-wave correlations between spin-polarized fermions in a multi-orbital q2D system with underlying $p$-wave interactions. The $s$-wave character of low-energy collisions in this regime manifests clearly in the scaling of the rf transfer rate. Furthermore, below-threshold bound states are found at energies predicted by dimensional-crossover theory for both $p$- and $s$-wave symmetry, corresponding to ground and excited orbital states, respectively. 
Our observations generalize the paradigm discovered recently in q1D systems \cite{Jackson:2023}. 

Future work could expand the two-body physics explored here to many-body, studying the dependence of the $p$- and emergent $s$-wave contact parameters on temperatures and magnetic field. A key question in this context will be understanding loss processes and orbital relaxation rates.
Another interesting direction would be the exploration of few-body states, including emergent $s$-wave halo dimers in the shallow ($B > B^*_s$) regime, or other exotic few-body states anticipated for q2D $p$-wave systems without orbital excitation     \cite{Lasinio:2008,DIncao:2008,Nishida:2013,Wang:2011}. 
Finally, the multi-orbital scattering paradigm introduced here can be extended to higher partial waves, e.g. to emergent q2D $p$-waves for strongly confined bosons or fermions near a free-space $d$-wave Feshbach resonance \cite{Cui2017,Yao2019,Fouche2019}. 

\begin{acknowledgments} 
We acknowledge K.\ Jackson for discussion and early contributions to this work. 
This research is financially supported by AFOSR FA9550-19-1-7044, AFOSR FA9550-19-1-0365, and NSERC. S.Z.\ is supported by HK GRF Grants No.\ 17306024 and No.\ 17313122, CRF Grants No.\ C6009-20G and No. C7012-21G, and a RGC Fellowship Award No.\ HKU RFS2223-7S03. 
\end{acknowledgments} 

\newpage
\bibliography{bib}


\newpage
\onecolumngrid

\newcommand{\PRLsep}{\noindent\makebox[\linewidth]{\resizebox{0.75\linewidth}{1.75pt}{$\bullet$}}\bigskip}

\appendix
\begin{center}
    \vspace{0.5cm}
    \PRLsep
    \vspace{0.5cm}
    \textbf{\Large Supplemental Material}
\end{center}

{\large \bf A. Anharmonic shifts} \\

Quasi-two-dimensional confinement in our experiment is provided by an optical standing wave, which is only harmonic to first order. This section estimates the spectral shifts in rf spectroscopy due to anharmonicity. 

In the strong confinement limit, the wavefunction is restricted to positions $x$ that are near the $x=0$ center of a given site. We can thus expand the lattice potential for $k_L x \ll 1$, where $k_L = 2 \pi / \lambda_L$, as 
\be V_L \sin^2(k_L x) \approx V_L (k_L x)^2 - \frac13 V_L (k_L x)^4 + \frac{2}{45} V_L (k_L x)^6 + \ldots \ee
The first term provides harmonic confinement. Comparing it to $V_\mathrm{ho} = \frac12 m \omega_\perp^2 x^2$, we recover the standard harmonic oscillation frequency,  
\be \label{eq:omega0}
\omega_\perp = ( {2 V_L k_L^2}/{m} )^{1/2} = 2 (V_L E_R/\hbar^2)^{1/2} = 2 \omega_R s^{1/2} \ee 
where $\omega_R = E_R/\hbar$ is the recoil energy in frequency units, and $s_L = V_L/E_R$ is the dimensionless lattice depth. We also note that the {\em single-particle} oscillator length $\ell_\perp \equiv (\hbar/m \omega_\perp)^{1/2}$ relates to the lattice depth as $\ell_\perp = k_L^{-1} s_L^{-1/4}$. 

For single particles, the first-order energy can be evaluated using the $\ket{n}$ harmonic-oscillator eigenstate: 
\be \begin{aligned}
\bra{n} V_L & \sin^2 (k_L \hat x) \ket{n} - \hbar \omega_\perp (n+1/2) \\
& = - \frac{V_L}{3} k_L^4 \bra{n} \hat x^4 \ket{n} + \frac{2 V_L}{45} k_L^6 \bra{n} \hat x^6 \ket{n} + \ldots \\ 
& = - \frac{E_R}{4} (2n^2 + 2n + 1) 
+\mathcal{O}\left(V_L^{-1/2} n^3 \right)
\end{aligned} \ee
We can then see that the energy gap between two successive single-particle bands has a leading correction
\be V^{(n+1)}_\mathrm{OL} - V^{(n)}_\mathrm{OL} \approx \hbar \omega_\perp - (n+1) E_R \,. \ee

Now consider two particles interacting with strong out-of-plane confinement. Neglecting in-plane confinement, the Hamiltonian is 
\be \hat H = \hat K_\mathrm{r} + \hat K_\mathrm{CM} + \hat V + \hat U_P \ee
where $\hat K_\mathrm{r}$ is kinetic energy of the relative motion, $\hat K_\mathrm{CM}$ is kinetic energy of the center-of-mass motion, $\hat V$ is the confining potential, and $U_P$ is the $p$-wave interaction. We can further expand the confining potential into its harmonic and anharmonic components: 
\be V(\hat x) = \hat V_\mathrm{ho} + \hat V^{(4)} + \mathcal O (\hat x^6) \ee
where \be \frac{ \hat V^{(4)} }{\hbar \omega_\perp} = - \frac{1}{6 \sqrt{V_L E_R}} V_L k_L^4 x^4 = \frac{-1}{6 s_L^{1/2} \ell_\perp^4} x^4  \equiv V_4  x^4 \label{eq:V4} \ee
where we have defined a $V_4$ with units of (length)$^{-4}$. 

When considering the two-particle problem, a complexity introduced by $V^{(4)}$ is the coupling of relative and CM motion (which are separable for harmonic confinement):  
\be \begin{aligned} 
\left.\frac{ \hat V^{(4)} }{\hbar \omega_\perp}\right|_A
& + \left.\frac{ \hat V^{(4)} }{\hbar \omega_\perp}\right|_B \\
& = V_4 \hat x_A^4 + V_4 \hat x_B^4 \\
&= V_4 (\hat x_\mathrm{CM} - \frac12 \hat x_\mathrm{r})^4 + V_4 (\hat x_\mathrm{CM} + \frac12 \hat x_\mathrm{r})^4 \\
&= \underbrace{\frac18 V_4}_{\equiv c_0} \hat x_\mathrm{r}^4 + \underbrace{3 V_4}_{\equiv c_1} \hat x_\mathrm{r}^2 \hat x_\mathrm{CM}^2 + 2 V_4 \hat x_\mathrm{CM}^4 
\end{aligned} \ee
where we have defined $c_0$ and $c_1$ for convenience. We can neglect shifts that depend only on the CM degrees of freedom, because it is not changed in rf spectroscopy; so, the $x_\mathrm{CM}^4$ term in this expression will be dropped. 
The anharmonic shift on a relative-motion state can then be found as 
\be \frac{ \langle  \hat V^{(4)}  \rangle }{\hbar \omega_\perp}
 = c_0 \left\langle \hat x_\mathrm{r}^4 \right\rangle + c_1 \left\langle \hat x_\mathrm{r}^2 \right\rangle \left\langle \hat x_\mathrm{CM}^2 \right\rangle \ee

Harmonic solutions considered in \cite{companion} include $\hat K_\mathrm{r} + \hat V_\mathrm{ho} + \hat U_P$; our task here is only to consider the shift due to $\hat V^{(4)}$. The observed shift is only the differential between the initial and final states:
\be \label{eq:DE4} 
\Delta E^{(4)} = \bra{f}\hat V^{(4)}\ket{f} - \bra{i} \hat V^{(4)} \ket{i} \ee
We will further approximate continuum states as weakly interacting along the strong confinement direction, and thus $\approx \ket{n_r}\ket{0_\mathrm{CM}}$. Thus,  
\be \frac{ \langle \hat V^{(4)}  \rangle }{\hbar \omega_\perp}
\approx c_0 \bra{n_\mathrm{r}} \hat x_\mathrm{r}^4 \ket{n_\mathrm{r}} + c_1 \bra{n_\mathrm{r}} x_\mathrm{r}^2 \ket{n_\mathrm{r}}
\bra{0_\mathrm{CM}} \hat x_\mathrm{CM}^2 \ket{0_\mathrm{CM}} \ee
%
These can be evaluated using harmonic-oscillator identities 
\be \begin{aligned}
\bra{n} \hat x^2 \ket{n} &= \frac12 a_\mathrm{ho}^2 (2n+1) \\ 
\bra{n} \hat x^4 \ket{n} &= \frac34 a_\mathrm{ho}^4 ( 2n^2 + 2n+1 ) \end{aligned} \ee 
where, in the case of relative motion, $a_\mathrm{ho} \to a_\perp=(2 \hbar/m \omega_\perp)^{1/2}$; and in the case of center-of-mass motion, $a_\mathrm{ho} \to a_\mathrm{\perp,CM} = a_\perp /2$. This gives
\be \begin{aligned} 
\frac{ \langle \hat V^{(4)}  \rangle }{\hbar \omega_\perp} &\approx 
c_0 \frac34 a_\perp^4 (  2n_\mathrm{r}^2 + 2n_\mathrm{r}+1 ) 
+c_1 \frac{1}{4} a_\perp^2 (2n_\mathrm{r}+1) \left(\frac{a_\perp}{2}\right)^2 \\
&= \frac{3}{32} V_4 a_\perp^4 (2n_\mathrm{r}^2 + 2n_\mathrm{r}+1) 
+\frac{3}{16} V_4  a_\perp^4 (2n_\mathrm{r}+1) 
\end{aligned} \ee 
We can write an expression in terms of lattice depth using
$V_4 = -s_L^{-1/2} \ell_\perp^{-4}/6$, 
and since $a_\perp/\ell_\perp = 2^{1/2}$, we have 
$V_4 a_\perp^4 = -\frac23 s_L^{-1/2}$. Also, $\hbar\omega_\perp = 2 s_L^{1/2} E_R$. 
Thus,
\be \begin{aligned} \label{DEinER}
\frac{\bra{n_r} \hat V^{(4)} \ket{n_r}}{E_R} &= -  
\frac18 (3 + 6 n_r + 2 n_r^2) \\
& = \left\{- \frac{3}{8}, -\frac{11}{8}, -\frac{23}{8}, -\frac{39}{8}, \ldots \right\} 
\end{aligned} \ee 
for $n_\mathrm{r} = \{ 0, 1, 2, 3, \ldots \}$ respectively. 
With $E_R/h = 8.6$\,kHz, the first four shifts are 
\{ $-3.2$\,kHz , $-12$\,kHz, $-25$\,kHz, and $-42$\,kHz\} respectively. 

For spectroscopy that projects $bb$ pairs onto the non-interacting $bc$ pair continuum, the ``step'' feature that appears at $2 \hbar \omega_\perp$ under harmonic confinement is reduced due to anharmonicity. Here we have $\ket{\psi_i} = \ket{n_r = 1, n_\mathrm{CM}=0}$ and $\ket{\psi_f} = \ket{n_r = 3, n_\mathrm{CM}=0}$. 
From Eqs.~\eqref{eq:DE4} and \eqref{DEinER}, the differential anharmonic shift is $\approx -3.5\,E_R$ or $\approx -30$\,kHz. This value is used to offset the step of the red dashed line in Fig.~1(e) of the main text. 

For spectroscopy that spin-flip $ab$ pairs in the scattering continuum into $bb$ pairs in a bound dimer state, we will make the approximation that the final state is a bound dimer, whose spatial extent is small; effectively, $\psi_f(x) \approx \delta(x)$. 
\be \Delta E^{(4)} \approx - \bra{0_\mathrm{CM}}\bra{\psi_i} \hat V^{(4)} \ket{0_\mathrm{CM}}\ket{\psi_i} \ee
In the $p$-wave case, $n_\mathrm{r} = 0$, and the anharmonic shift is $\Delta E^{(4)}/h \approx +3.2$\,kHz. For the $s$-wave case, $n_\mathrm{r} = 1$, and the anharmonic shift is $\Delta E^{(4)}/h \approx +12$\,kHz. These values are shown in the inset to Fig.~2 in the main text. 

\bigskip 
{\large \bf B. Dimer lineshape model} \\

The fitting model includes a Frank-Condon factor $F_\mathrm{fi}$ and a two-body thermal distribution $P(T, E_k)$, where $T$ is the temperature and $E_k + U = \hbar \omega$. The number of atoms lost from the trap is
\be \delta N(\omega_\mathrm{rf} ) = A P(T,E_k) F_\mathrm{fi}\,, \ee 
where $A$ is a fit parameter. We assume that incident particles have a 2D Maxwell-Boltzmann distribution of relative energy, a semi-classical collision rate proportional to $E_k^{1/2}$, and that the Frank-Condon factor scales with relative energy and binding energy like $F_\mathrm{fi}\propto E_k/U_p^2$ for $p$-wave dimers, and $F_\mathrm{fi}\propto 1/U_s$ for $s$-wave dimers. We ignore interaction effects. The lineshapes are then
\begin{align} N(\omega) &= N_{\mathrm{bg}} - A\,\frac{(U_p+\omega)^{3/2}}{U_p^2} e^{-(U_p +\omega)/k_B T}\,, \\
N(\omega) &= N_{\mathrm{bg}} - A\,\frac{(U_s+\omega)^{1/2}}{U_s} e^{-(U_s +\omega)/k_B T}\,, \end{align}
for $p$- and $s$-wave dimers, respectively, where the fit parameter $N_\mathrm{bg}$ is determined by the average number of atoms in the absence of loss. The thermal distribution and Frank-Condon factor make the lineshape asymmetric, where the fit $U_{\{p,s\}}$ is near the edge of the loss feature which corresponds to zero relative motion of the free particles. The lineshape fits give a binding energy that is on average $10\,$kHz shifted from a Gaussian fit to peak loss.

Inhomogeneity in the lattice depth across the ensemble of q2D systems broadens the expected dimer energy measurement, since the binding energy depends on the depth of the confining potential. The lattice-depth distribution in the ensemble is estimated from bandmapping frequency modulation spectra of the lattice, which transfers atoms from the ground to the excited band. A variation of lattice depth of $7\%$ and $15\%$ are observed in the spectra for $x$- and $z$-confinement geometries, respectively. Loss features are fit to a sum of lineshapes for each depth, weighted by the lattice-depth distribution. This has a small $1$-$2\,$kHz shift on estimates of $U_p$, but a larger $5$-$20\,$kHz shift on $U_s$.

In general, the dimer-association signal for the $s$-wave dimers is smaller than $p$-wave dimers, since less than half of the atoms are in the orbitally excited motional state ($f_e<0.5$) required to pair with a ground-motional-state atom for dimer association. Furthermore, lattice inhomogeneity has a larger effect on the $s$-wave association features, since the orbitally excited initial state depends more strongly on the lattice depth. The result is a wider and smaller association feature. This reduces the precision for which the $U_s$ can be determined compared to $U_p$. 

The best-fit parameters and uncertainty estimation for each rf spectra are obtained by using the Monte Carlo Bootstrap method. Each spectra is randomly sampled $N$ times from the $N$ data points with replacement, and this set of points is fit to the appropriate lineshape. We repeat this 5000 times for each spectra, excluding datasets when the best-fit parameter distributions are non-Gaussian or heavily skewed. The remaining spectra use the median of the parameter distributions as the overall best fit, with one-sigma confidence intervals as the uncertainties. This statistical method illuminates situations when the lineshape edges, associated with the binding energies, are constrained by only a few data points.

The use of a Fermi-Dirac distribution in place of a Boltzmann as the thermal distribution introduces only a small shift in best-fit binding energy, which is smaller than the widths of the fit-parameter distributions.

\end{document}